\documentclass[a4paper]{article}
\usepackage{INTERSPEECH2021}
\usepackage{amsmath,graphicx}
\usepackage{arydshln}

\usepackage{comment}
\usepackage[table]{xcolor}
\usepackage{hyperref}

\hypersetup{colorlinks=true}

\def\sh{\cellcolor{gray!20}}

\definecolor{my_green}{HTML}{228B22}

\title{SpeakerStew: Scaling to Many Languages with a Triaged Multilingual Text-Dependent and Text-Independent Speaker Verification System}
\name{Roza Chojnacka$^*$ \qquad Jason Pelecanos$^*$ \qquad Quan Wang \qquad Ignacio Lopez Moreno\thanks{* Equal contribution.}}
{
\address{Google LLC, USA}
\email{\{\href{mailto:roza@google.com}{roza},\href{mailto:pelecanos@google.com}{pelecanos},\href{mailto:quanw@google.com}{quanw},\href{mailto:elnota@google.com}{elnota}\}@google.com}}

\begin{document}

\interfootnotelinepenalty=10000
\maketitle
\begin{abstract} 
In this paper, we describe \emph{SpeakerStew} -- a hybrid system to perform speaker verification on 46 languages. Two core ideas were explored in this system: (1) Pooling training data of different languages together for multilingual generalization and reducing development cycles; (2) A novel triage mechanism between text-dependent and text-independent models to reduce runtime cost and expected latency.
To the best of our knowledge, this is the first study of speaker verification systems at the scale of 46 languages.
The problem is framed from the perspective of using a smart speaker device with interactions consisting of a wake-up keyword (text-dependent) followed by a speech query (text-independent).
Experimental evidence suggests that training on multiple languages can generalize to unseen varieties while maintaining performance on seen varieties. We also found that it can reduce computational requirements for training models by an order of magnitude. Furthermore, during model inference on English data, we observe that leveraging a triage framework can reduce the number of calls to the more computationally expensive text-independent system by 73\% (and reduce latency by 59\%) while maintaining an EER no worse than the text-independent setup.



\end{abstract}

\noindent\textbf{Index Terms}: SpeakerStew, speaker recognition, multilingual, cross-lingual, hybrid system, triage

\section{Introduction}    
\label{sec:introduction}

Speech applications are becoming more and more pervasive with their popularisation in cell phones, cars and smart speakers. With increased adoption there is a demand for speech applications in more languages. This introduces challenges such as creating speech resources for the unsupported languages, building new models for the added languages and maintaining them while in production. This entails significant human and computational effort. Many speech tasks naturally call for a single model to represent all languages.

We explore this problem from the perspective of a speaker verification task on smart speaker devices, although similar principles could be applied to other select scenarios. For a new language to be added, speech data including audio and speaker labels from the language of interest must be available. A language specific model is then trained on this dataset and operating thresholds are determined for the end application considered. For the dataset studied in this paper, there are 46 languages of interest. Deploying models across all 46 languages involves significant effort. This becomes more challenging when we need to maintain a lightweight text-dependent (TD) model and a larger text-independent (TI) model. In this kind of setup, a natural question comes up whether it's possible to perform well on seen languages and generalize well to unseen languages while reducing computational requirements.

Researchers have studied various speech data augmentation techniques to help speaker verification systems better generalize. Such approaches include speed perturbation~\cite{ko2015audio}, room simulation~\cite{kim2017generation,ko_2017,snyder_2018_1}, SpecAugment~\cite{park2019specaugment,faisal2019specaugment,wang2020investigation}, and speech synthesis~\cite{huang2020synth2aug}. However, these techniques are mostly known to make speaker verification more robust to noise, channel effects and related domain mismatch, instead of language mismatch.

Studies have covered language mismatched system training and cross-language speaker verification trials. NIST~\cite{greenberg_2020_1, nist_1}, a standards organization, supported its first non-English evaluation in 2000 (AHUMADA corpus~\cite{ortega_2000_1}) followed by its first cross-language trial tasks in 2005/2006. Lu~\cite{lu_2009} improved evaluation results on non-English trials of a model trained only on English data. This was achieved by removing language factors from a joint factor analysis model. There was also a study~\cite{misra_2014} to extract speaker features using deep learning. The CNN-TDNN model was trained on English data and evaluated on the Chinese and Uyghur languages. It outperformed the baseline i-vector model by a large margin, particularly for language mismatch between enrollment and test. Another paper~\cite{kumar_2009} assessed a neural network based multilingual model trained on English and 4 Indian languages. More recently,~\cite{xia2019_1} examined adversarial training techniques and~\cite{thienpondt2020_1} explored the use of language-dependent score normalization. In~\cite{perrachione_2017} the authors mentioned that listeners identify voices in their native language more accurately than voices in an unknown, foreign language. 

There have been multiple works examining the combination of text-dependent and text-independent information. For the 2001 NIST text-independent task~\cite{nist_1}, Sturim~\cite{sturim_2002_1} selected commonly spoken keywords to model separately and then combined the result with a text-independent system. In the commercial space, Telstra, KAZ and Nuance delivered a speaker verification solution to Centrelink, the Australian Government Social benefits arm. It involved combining the results of multiple systems based on text-dependent and text-independent components~\cite{top_2009_1, summerfield_2008_1}.

Past papers have explored combining text-dependent and text-independent components generally using linear system combination to optimize for performance. In this paper we explore a novel triage mechanism for system combination such that we not only maintain close to optimal performance but can also reduce the overall computational burden and response time in returning the result. Additionally, we disseminate results studying how performance varies across high and low resource languages. Recently,~\cite{toshniwal2018multilingual} and~\cite{pratap2020massively} built a single acoustic model for multiple languages with the aim of improving automatic speech recognition (ASR) performance on low-resource languages. In our SpeakerStew system, we aim to improve speaker recognition performance such that it better generalizes across languages, and even languages without training data.


The rest of this paper is organized as follows. Section~\ref{sec:system} describes the system which includes the proposed triage mechanism, Section~\ref{sec:experiments} details the experimental setup and corresponding results, and Section~\ref{sec:conclusions} wraps up with the conclusions.

\vspace{-0.2cm}
\section{System description}
\label{sec:system}


In this section we discuss the 3 main components of the system; the lightweight TD system, the larger TI system, and a triage mechanism to combine the two information sources. These systems are designed to process a wake-up keyword (such as \textit{``Ok Google"}) followed by a query (for example, \textit{``What is the weather tomorrow?"}).

\subsection{TD system} 

The TD system has a small memory footprint (235k parameters). It is trained to perform verification for two sets of keywords: \textit{``Ok Google"} and \textit{``Hey Google"}. The same keywords are used across languages but pronunciation may be different. This system is based on the TD work in~\cite{wan_2018_1} and we discuss key details here.   

Firstly, for the TD system to work, we need to extract the relevant segment of speech containing only the keyword. For this purpose we used a multilingual keyword detector trained on all available languages and more details are available here~\cite{park_2020}. Once the segment containing the keyword is identified, this audio is parameterised into frames of 25ms width and 10ms shift. We extract 40-dimensional log Mel filter bank energy features for each frame and stack 2 subsequent frames together. To improve robustness of the model, we apply data augmentation techniques involving noise sources combined with room simulation effects ~\cite{lippmann_1987, ko_2017, kim_2017}. The resulting features are globally normalized to improve neural network training. 

Our neural network consists of 3 LSTM layers with projection and a linear transform layer. Each LSTM layer has 128 memory cells, and is followed by a projection layer of size 64 with a tanh activation~\cite{sak_2018_1}. The 64-dimensional linear transform layer is applied to the output of the last projected LSTM layer, which is followed by L2-normalization to produce the speaker embedding. 

We use the cosine similarity to compare enrollment utterances with evaluation utterances. We note that each speaker can have multiple enrollment utterances. Enrollment embeddings (which are already L2 normalized) are averaged for each speaker and then L2 normalized again. We used Generalized end-to-end (GE2E) contrastive loss~\cite{wan_2018_1} for training.

\subsection{TI system}

In contrast to the TD system, the TI system involves a larger model with 1.3M parameters. It uses both the keyword and speech query audio to recognize the speaker. For this TI task, it is not only that the pronunciation is different but the language can also differ. The system is based on the TI model proposed in~\cite{wan_2018_1} and is the same as the TD system except that it is a larger model and the training criterion is different. For the TD setup we use 384 memory cells with a projection size of 128. The final output speaker embedding is also of 128 dimensions. The training criterion uses the GE2E eXtended Set (XS) softmax loss~\cite{pelecanos2021_1}.


\subsection{System triage}  
\label{sec:triage}

\begin{figure}[htb]
 \vspace{-0.3cm}
 \centering
 \centerline{\includegraphics[width=8.0cm]{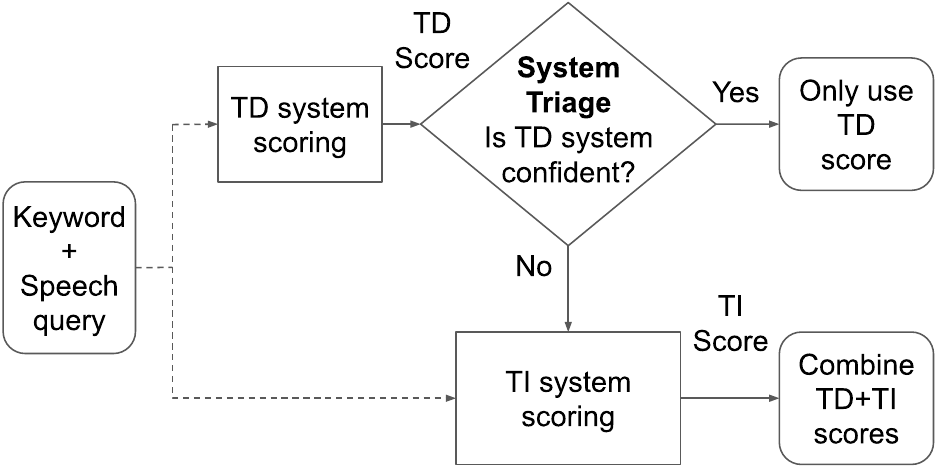}}
 \vspace{-0.3cm}
 \caption{Triage setup for speaker verification.}
\label{fig:system_triage}
\vspace{-0.5cm}
\end{figure}

In this section we discuss the use of system triage as an approach for maintaining speaker recognition performance while reducing latency and computation. In this work, we use a small-footprint, text-dependent speaker recognition system and a larger text-independent system. It is proposed that when the TD system is relatively confident, only the score for the TD system is used. When the TD system is not confident, scores from both the TD and TI systems are combined. With this approach, computation can be reduced. Additionally, for confident TD decisions, this information can be passed to other components sooner without having to wait for the entire query to be spoken. For example, language models can benefit from knowing the speaker identity in advance and fetch contextual information such as contacts, playlists or search history for this speaker before processing the query~\cite{aleksic2015bringing}. Figure~\ref{fig:system_triage} illustrates the triage mechanism.


\section{Experimental results}
\label{sec:experiments}

In this section we introduce the experiment setup, the data used in training and evaluation, and experiments using monolingual and multilingual models followed by a system triage analysis.

\subsection{Training and evaluation data} 

For training we use vendor collected speech data which contain recordings with \textit{``OK Google"} and \textit{``Hey Google"} keywords followed by a speech query. The data set covers 46 languages (63 dialects)\footnote{A list of 46 languages: Arabic, Bulgarian, Bengali (India), Mandarin Chinese (Simplified), Mandarin Chinese (Traditional), Czech, Danish, German, Greek, English, Spanish, Finnish, Filipino, French, Galician, Gujarati (India), Hebrew, Hindi, Croatian, Hungarian, Indonesian, Italian, Japanese, Kannada (India), Korean, Lithuanian, Malayalam (India), Marathi (India), Malay, Norwegian, Dutch, Polish, Portuguese, Romanian, Russian, Slovak, Serbian, Swedish, Tamil (India), Telugu (India), Thai, Turkish, Ukrainian, Urdu (India), Vietnamese, Cantonese.}. The data are not evenly distributed across languages. For example, training data range from 39k utterances to 1.9M training utterances. Table~\ref{tab:data_stats} shows the number of utterances and speakers for selected languages in training and evaluation. The first group (de, en, es, fr, ja, ko, pt) represents \textit{seen languages} which are used in the training of the \textit{Multilingual-12} model used later. The second group contains \textit{unseen languages} which are languages not used in the \textit{Multilingual-12} model. Datasets of different languages are combined with the MultiReader approach during training~\cite{wan_2018_1}. Evidence from speech recognition also indicates combining more training data improves speech models in general~\cite{chan2021speechstew}.

\subsection{Experimental setup} 
\label{sec:experimental_setup}


We trained 14 TD models and 14 TI models. For each of the TD/TI model sets, 12 are monolingual and trained only on data from the corresponding language, and the other two are multilingual models. The \textit{Multilingual-46} model was trained on all available data from 46 languages and the \textit{Multilingual-12} model was trained on 12 languages\footnote{Multilingual-12 training data languages: Danish (da), German (de), English (en), Spanish (es), French (fr), Italian (it), Japanese (ja), Korean (ko), Norwegian (nb), Dutch (nl), Portuguese (pt), Swedish (sv).}. We used the same structure and parameters for both the monolingual and multilingual models.

\begin{table}[htb]
\setlength{\tabcolsep}{3.9pt}
\centering
  \caption{Number of speakers (Spk), utterances (Utt), same-speaker (Same) and different-speaker (Diff) trials. Data is partitioned into training and evaluation data across languages\protect\footnotemark. The first set of languages represents a sampling of 7 of the 12 languages used in training the \textit{Multilingual-12} model. The second set of languages represent a sampling of the remaining 34 languages not seen by the \textit{Multilingual-12} model. The last 2 rows show the total number of speakers and utterances used to train the \textit{Multilingual-12} and the \textit{Multilingual-46} models.}
  \vspace{-0.1cm}
  \label{tab:data_stats}
\begin{tabular}{l | r r | r r r r}  
                          & \multicolumn{2}{c|}{\textbf{Training}} & \multicolumn{4}{c}{\textbf{Evaluation}} \\
  \textbf{Language} & \textbf{Spk} & \textbf{Utt} & \textbf{Spk} & \textbf{Utt} & \textbf{Same} & \textbf{Diff} \\
  \textbf{} & \textbf{[k]} & \textbf{[k]} & \textbf{[k]} & \textbf{[k]} & \textbf{[k]} & \textbf{[k]} \\ \hline
  
  German (de) & 3.4 & 908 & 1.5 & 255 & 192 & 188\\
  English (en)   &  38.0 &  4531 &  0.2 &  13 & 12 & 189 \\
  Spanish (es) & 18.0 & 1940 & 2.6 & 164 & 159 & 191\\
  French (fr) & 14.7 & 905 & 0.4 & 34 & 57 & 189 \\
  Japanese (ja) & 14.3 & 1308 & 0.4 & 29 & 21 & 116 \\
  Korean (ko) & 10.1 & 827 & 0.5 & 100 & 27 & 195 \\
  Portuguese (pt) & 14.3 & 927 & 1.5 & 113 & 114 & 196 \\ \hline
  Arabic (ar) & 4.6  & 463  & 1.5 & 115 & 102 & 174 \\
  Mandarin (cmn) & 3.6 & 310 & 0.6 & 54 & 33 & 110 \\
  Polish (pl) & 0.1 &  39 & 0.025 & 10  & 9 & 18 \\ 
  Russian (ru) & 5.9 & 398 & 0.4 & 33 & 31 & 192\\
  Vietnamese (vi) & 2.8  & 224  &  0.3 & 41 & 39 & 185 \\
  Malay (ms) & 2.1 & 168 & 0.4 & 33 & 26 & 150 \\ \hline
  Multilingual-12 & 126.0 & 12619 & - & - & - & - \\ 
  Multilingual-46 & 196.0 & 20618  & - & - & - & - \\ 
\end{tabular}
  \vspace{-0.5cm}
\end{table}
\footnotetext{For English, training utilizes multiple varieties of English while evaluation is performed on United States (US) English only.}

\subsection{Monolingual model performance}  

The evaluation results for different languages are shown in Table~\ref{tab:monolingual_models}. They show the effectiveness of replacing one monolingual model with another. Unsurprisingly, the best results are observed for the cases where the training and evaluation languages are the same. Some models generalize better than others. For example, in the absence of the Spanish (es) model, the Portuguese (pt) model would be the next best option on Spanish language data. It is reassuring that they are related languages.

The other observation is that the TI models are significantly better than the TD models. One of the reasons is that the TI model has a larger capacity (1.3M parameters in TI vs 235k parameters in TD). Another reason is that the TI system can utilize more audio than the TD system. The TI system uses both the keyword and the query as evidence to determine similarity, whereas the TD system uses only the keyword.

\begin{table}[htb]
\setlength{\tabcolsep}{4.8pt}
\centering
  \caption{Evaluation of monolingual models across languages. The columns represent different languages that were evaluated. The rows describe the type of model (TD/TI) and the language the model was trained on. The shaded cells represent models that have not seen training data with the same language as the evaluation data set. (Performance measured as \% EER.)}
  \vspace{-0.1cm}
  \label{tab:monolingual_models}
  \begin{tabular}{l | c  c  c  c  c  c  c }
    \textbf{TD/TI}  & \multicolumn{7}{c} {\textbf{Performance on different languages}} \\
  \textbf{System} & \textbf{de} & \textbf{en} & \textbf{es} & \textbf{fr}   & \textbf{ja} & \textbf{ko}  & \textbf{pt}  \\ \hline
  TD-de  &  1.27    &   \sh{5.54}   &    \sh{1.86}  &  \sh{3.04}    & \sh{4.32}     &   \sh{3.61}   &  \sh{3.09}   \\
  TI-de  & 0.61 & \sh{1.94} & \sh{0.72} & \sh{1.31} & \sh{1.22} & \sh{1.32} & \sh{1.01}   \\ \hline
  TD-en    &   \sh{2.04}   &  2.64    &    \sh{1.66}  &   \sh{2.05}   &    \sh{3.14}  &     \sh{3.05} &    \sh{1.81}   \\
  TI-en    & \sh{0.96} & 1.13 & \sh{0.82} & \sh{1.17} & \sh{0.69} & \sh{1.24} & \sh{0.86} \\ \hline
  TD-es   &     \sh{1.80} &  \sh{3.20}    &   0.64   & \sh{1.88}     &    \sh{2.83}  &    \sh{2.69}  & \sh{1.50}  \\
 TI-es   &  \sh{1.01} & \sh{1.43} & 0.54 & \sh{1.22} & \sh{0.69} & \sh{1.06} & \sh{0.83} \\ \hline
  TD-fr     &    \sh{2.23}  &   \sh{3.51}   &    \sh{1.98}  &    1.29  &   \sh{3.40}   &    \sh{2.74}  &    \sh{2.12}  \\
  TI-fr  & \sh{0.94} & \sh{1.37} & \sh{0.65} & 0.92 & \sh{0.64} & \sh{0.90} & \sh{0.81}\\ \hline
  TD-ja     &    \sh{2.95}  &    \sh{4.55}  & \sh{2.57}     &  \sh{2.81}    &    1.12  &     \sh{3.26} &    \sh{2.56}  \\
  TI-ja    & \sh{1.18} & \sh{1.54} & \sh{0.75} & \sh{1.14} & 0.32 & \sh{1.02} & \sh{0.88}\\ \hline
  TD-ko     &    \sh{2.85}  &  \sh{4.58}    &    \sh{2.38}  & \sh{2.98}     &    \sh{3.56}  &    1.05 &   \sh{2.43}   \\
  TI-ko     & \sh{1.04} & \sh{1.65} & \sh{0.78} & \sh{1.14} & \sh{0.81} & 0.39 & \sh{0.86} \\ \hline
  TD-pt     &    \sh{2.05}  &   \sh{3.39}   &  \sh{1.46}    &  \sh{2.06}    &     \sh{3.55} &   \sh{2.58}    &  0.93    \\
  TI-pt     &    \sh{0.82} & \sh{1.21} & \sh{0.52} & \sh{0.94} & \sh{0.62} & \sh{0.98} & 0.48\\  \hline
  \end{tabular}
  \vspace{-0.5cm}
\end{table}

\subsection{Multilingual model performance}  

Table~\ref{tab:multilingual_models} compares the results across monolingual and multilingual trained systems. It is observed that the multilingual models can reach or even exceed the performance of monolingual models for both seen and unseen languages. It is observed that for languages with very few training speakers, a multilingual model can be of significant benefit. Polish is a language with only about 100 training data speakers. The performance on Polish can be greatly improved by using multilingual models over the monolingual counterparts.

\begin{table*}[htb]
\centering
  \caption{Table comparing the performance of (i) TD models (ii) TI models and (iii) the optimal linear score combination of the corresponding TD and TI systems. The columns represent the different languages that were evaluated. The left side of the table contains evaluation results for higher-resource languages that were used to train the \textit{multilingual-12} models. The right side represents lower-resource languages that the \textit{multilingual-12} models have not seen and the corresponding results are shaded to indicate the unseen language condition. The \textit{multilingual-46} models have seen all languages. Monolingual models were trained using data from one language only (the same as evaluation language). Performance measured as \% EER.}
  \vspace{-0.3cm}
  \label{tab:multilingual_models}
\begin{tabular}{l | c  c  c  c  c  c  c | c  c  c  c  c  c }
      & \multicolumn{7}{c|}{\textbf{Languages in \textit{multilingual-12} model}}  & \multicolumn{6}{c}{\textbf{Languages \underline{not} in \textit{multilingual-12} model}} \\
  \textbf{System} & \textbf{de} & \textbf{en} & \textbf{es} & \textbf{fr}   & \textbf{ja} & \textbf{ko}  & \textbf{pt} & \textbf{ar} &\textbf{cmn} & \textbf{pl} & \textbf{ru} & \textbf{vi} & \textbf{ms} \\ \hline
  TD-monolingual       & 1.27 & 2.64 & \textbf{0.64} & 1.29 & \textbf{1.12} &\textbf{ 1.05} & \textbf{0.93} & 2.26 & 1.96 & 17.89 & 2.10 & 3.15 & 3.72 \\
  TD-multilingual-12              & \textbf{1.09} & \textbf{2.47} & 0.77 & 1.28 & 1.42 & 1.48 & 1.11 & \sh{1.60} & \sh{1.82} & \sh{\textbf{1.01}} & \sh{1.40} & \sh{2.48} & \sh{1.76} \\
  TD-multilingual-46              & 1.16 & 2.53 & 0.80 & \textbf{1.24} & 1.64 & 1.62 & 1.14 & \textbf{1.53} & \textbf{1.27} & 1.03 & \textbf{1.34} & \textbf{1.93} & \textbf{1.59} \\ \hline
  TI-monolingual       & 0.61 & 1.13 & 0.54 & 0.92 & 0.32 & \textbf{0.39} & 0.48 & 1.05 & 0.60 & 11.78 & 0.87 & 1.39 & 1.58 \\
  TI-multilingual-12              & \textbf{0.44} & 0.91 & 0.29 & \textbf{0.65} & \textbf{0.24} & 0.40 & \textbf{0.35} & \sh{0.77} & \sh{0.42} & \sh{0.91} & \sh{\textbf{0.64}} & \sh{0.56} & \sh{\textbf{0.79}} \\
  TI-multilingual-46              & 0.49 & \textbf{0.84} & \textbf{0.24} & 0.66 & 0.28 & 0.48 & \textbf{0.35} & \textbf{0.65} & \textbf{0.38} & \textbf{0.50} & 0.66 & \textbf{0.51} & 0.80 \\ \hline
  
  Comb-monolingual     & 0.47 & 0.98 & 0.29 & 0.69 & 0.24 & \textcolor{my_green}{{\textbf{0.25}}} & 0.34 & 0.78 & 0.36 & 8.47 & 0.64 & 0.81 & 1.00 \\
 
 Comb-multilingual-12            & {\textcolor{my_green}{\textbf{0.40}}} & 0.86 &\textcolor{my_green}{{\textbf{0.22}}} & \textcolor{my_green}{{\textbf{0.57}}} & \textcolor{my_green}{{\textbf{0.22}}} & 0.33 & \textcolor{my_green}{{\textbf{0.30}}} & \sh{0.64} & \sh{0.34} & \sh{0.58} & \textcolor{my_green}{\sh{\textbf{{0.62}}}} & \sh{0.53} & \textcolor{my_green}{\sh{\textbf{{0.78}}}} \\
 
 Comb-multilingual-46            & 0.45 & \textcolor{my_green}{{\textbf{0.83}}} & \textcolor{my_green}{{\textbf{0.22}}} & 0.60 & 0.24 & 0.39 & 0.31 & \textcolor{my_green}{{\textbf{0.57}}} & \textcolor{my_green}{{\textbf{0.28}}} & \textcolor{my_green}{{\textbf{0.48}}} & 0.64 & \textcolor{my_green}{{\textbf{0.48}}} & \textcolor{my_green}{{\textbf{0.78}}} \\ \hline
\end{tabular}
\vspace{-0.5cm}
\end{table*}



\subsection{Linear score combination}   

It's beneficial to combine TD and TI models - these results are shown in the last set of rows of Table~\ref{tab:multilingual_models}. The performance numbers shown here are based on finding an optimal linear combination weight (using a linear sweep) of the corresponding TD/TI systems for each combination result in the table.

An observation is that for the multilingual TD/TI linear score combination results, the \textit{Comb-multilingual-12} (\textit{12L}) system is comparable to the \textit{Comb-multilingual-46} (\textit{46L}) system for languages seen by the \textit{12L} model. For languages not seen by the \textit{12L} model, there is a slight performance improvement with the \textit{46L} setup. For most practical purposes, the \textit{12L} model is sufficient to capture the performance benefit without resorting to requiring data from 46 languages. Additionally, the \textit{12L} model was not trained on tonal languages yet generalizes well to Mandarin (cmn) and Vietnamese (vi) which are tonal languages.

\subsection{Triage score combination}    
In this section we examine how system triage can maintain speaker recognition performance while improving average decision response time (latency) and computational requirements.

In our triage implementation, we use the TD system score when the TD system is confident. When it is not confident we generate the TI system score and combine it with the TD system result. The two scores are combined based on  using the same optimal combination weights for the corresponding language/model in Table~\ref{tab:multilingual_models}. In determining whether the TD system is confident, we check whether the TD score is inside or outside a range defined by upper and lower score thresholds. If it is outside the selected score bounds, then the TD system is regarded as confident. Otherwise the TD system is not confident.

\begin{figure}[htb]
 \centering
 \centerline{\includegraphics[trim=0cm 3cm 0cm 3.0cm,clip=true,width=8.0cm]{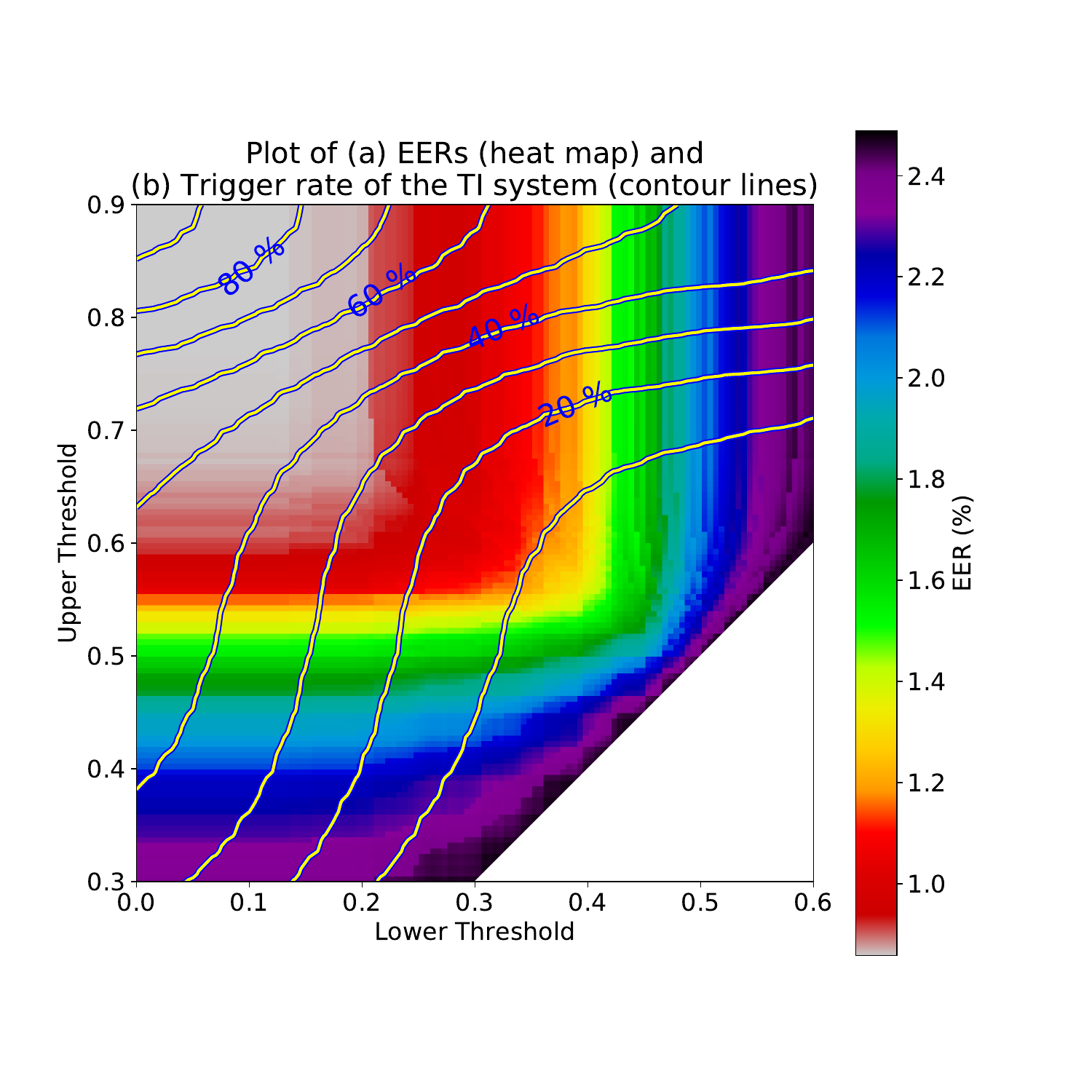}}
 \vspace{-0.3cm}
 \caption{Triage results showing the EER (heat map) and the TI trigger rate (contour lines) as a function of the TD upper and lower score thresholds for US English evaluation data and the \textit{TD/TI-multilingual-12} models. Same-speaker class probability is 50\%. }
\label{fig:triaged_eer}
\vspace{-0.3cm}
\end{figure}

Figure~\ref{fig:triaged_eer} consists of two parts; (i) an EER heat map and (ii) a contour plot overlay showing the TI trigger rate (\textit{i.e.} how often the TI system is needed). The heat map presents the EER as a function of the lower (x-axis) and upper (y-axis) cut-off thresholds of the TD system. The upper left region of the plot sustains better performance while moving away from this region increases the EER. This heat map needs to be considered along with the contour plot overlay which describes the percentage of trials that use both the TI and TD systems (\textit{i.e.} the TI trigger rate). For the \textit{multilingual-12} model, the English TD EER is 2.47\% and the TI EER is 0.91\% while the always combined result is 0.86\% (Table ~\ref{tab:multilingual_models}). If we set lower and upper thresholds of 0.23 and 0.65 respectively, we can maintain TI EER performance while only needing the TI system for 27\% of trials (Figure~\ref{fig:triaged_eer}), a 73\% reduction. In our application of interest, the average duration of the keyword portion of speech is 0.7s, and the query portion is 3s on average. At 27\% TI trigger rate, assuming an immediate system response, the expected time to verification completion is reduced from 3.7s (the utterance duration) to 1.5s. This is a saving of 2.2s (59\% reduction). 

\begin{figure}[htb]
 \centering
 \centerline{\includegraphics[width=7.5cm]{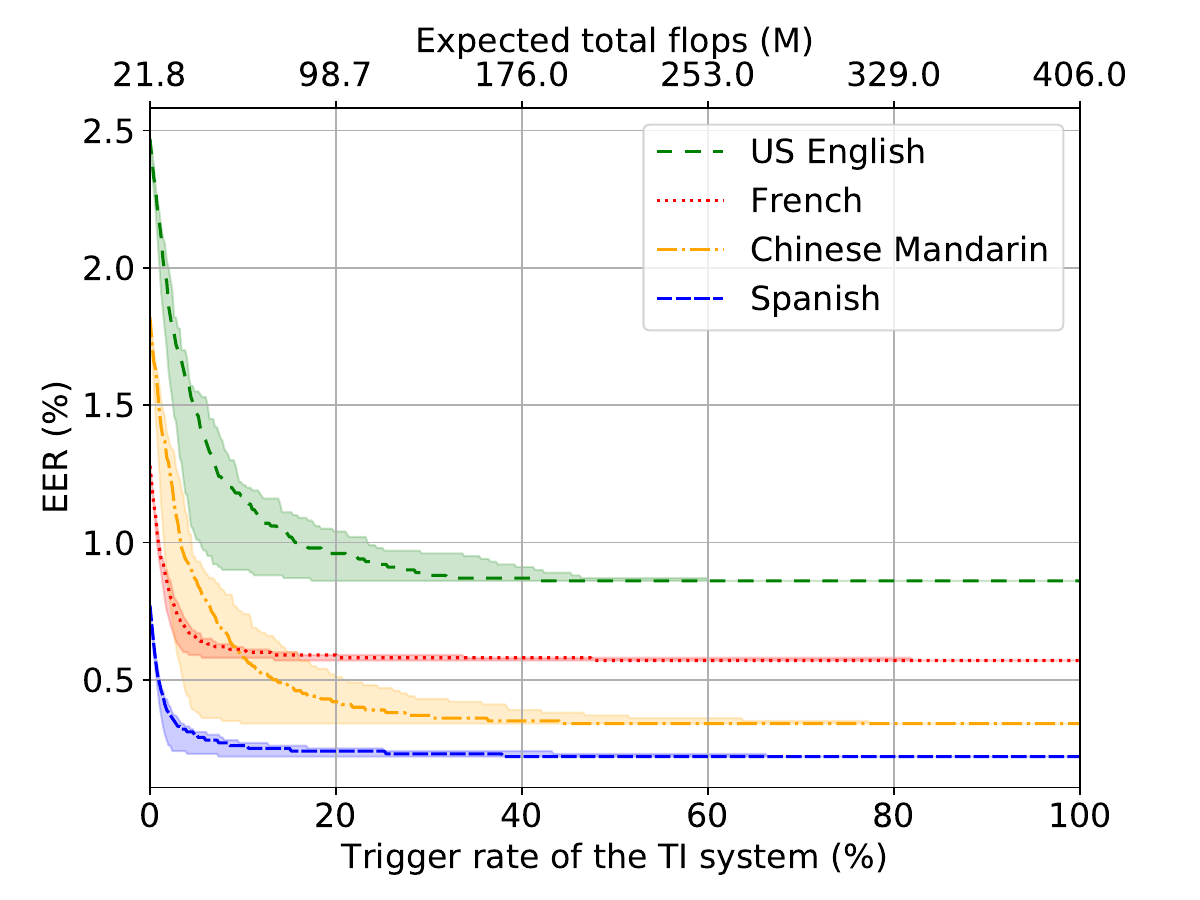}} 
 \vspace{-0.3cm}
 \caption{Plot of EER as a function of the TI trigger rate as well as the expectation of the total number of floating point operations (flops) for the \textit{multilingual-12} system. For each language a dashed error curve is shown for the case with equal same/different speaker class prior probabilities. The corresponding shaded area reveals the EER bounds (i.e. best/worst case scenarios) determined when the same-speaker-trial class probability is either close to 0\% or 100\%.
}
\label{fig:triaged_eer_vs_efficiency}
\end{figure}

There are two factors to consider when setting the thresholds. The first is that the contour overlay and the suitable thresholds chosen are affected by the prior probabilities of the same-speaker and different-speaker classes. In Figure~\ref{fig:triaged_eer}, there is a 50\% same-speaker-trial probability. The second consideration is if this approach is to operate with the same thresholds across all languages. If so, it may require a larger range for the lower and upper cutoff thresholds and would reduce efficiency. Figure~\ref{fig:triaged_eer_vs_efficiency} examines this first factor by evaluating how performance can be affected by class probabilities. For clarity, four languages are shown. The dashed lines show the result for 50\% same-speaker-trial probability while the shaded areas show the best/worst case scenarios by assessing performance at 0\% and 100\% same-speaker-trial probabilities. For this framework, the figure shows that worst case performance is not much worse than the equal class probability scenario.




\section{Conclusions}    
\label{sec:conclusions}

Our experiments provide evidence toward the hypothesis that a
speaker recognition model trained on data from different languages  
generalizes better to unseen languages than monolingual models. This applies to both text-dependent (where the keyword is pronounced differently in each language) and text-independent models. In scaling up to many languages, an opportunity is presented to improve efficiency~\cite{Strubell_Ganesh_McCallum_2020}. During training, bundling languages into a single model can reduce computational requirements by an order of magnitude. At runtime, triage can be implemented to reduce the calls to a larger text-independent model without significantly degrading overall performance, thus reducing latency and computational cost.




\clearpage
\vfill\pagebreak
\clearpage

\bibliographystyle{IEEEtran}
\bibliography{refs}

\end{document}